%% file: main.tex
\documentclass[conference]{IEEEtran}

\input{commands}

\ifCLASSINFOpdf
\else
\fi
\begin{document}

\title{From Investigation Failures to Reliable SOC Agents: Understanding and Improving LLM-Based Alert Triage}


\author{\IEEEauthorblockN{Saimon Amanuel Tsegai}
	\IEEEauthorblockA{Virginia Tech\\
		saimon.tsegai@vt.edu}
	\and
    \IEEEauthorblockN{Alex Kantchelian }
	\IEEEauthorblockA{Google\\
		akant@google.com}
    \and
    \IEEEauthorblockN{Danfeng (Daphne) Yao}
	\IEEEauthorblockA{North Carolina State University\\
		daphne\_yao@ncsu.edu}
    \and
	\IEEEauthorblockN{Peng Gao}
	\IEEEauthorblockA{Virginia Tech\\
		penggao@vt.edu}
}



\maketitle

\begin{abstract}
    \input{sections/abstract}

\end{abstract}


\IEEEpeerreviewmaketitle


\input{sections/introduction}


\input{sections/background}


\input{sections/benchmark}


\input{sections/systematic_study}


\input{sections/system_design}


\input{sections/evaluation-new}


\input{sections/discussion}


\input{sections/conclusion}




\section*{Ethics Considerations}
                                
We evaluate AIDA using publicly available datasets in isolated replay environments. The study does not involve human subjects or private user data, and the evaluation does not interact with external or operational systems. The main risks arise in real-world deployment and from dual use. AIDA may process sensitive security telemetry and may make incorrect decisions when evidence is incomplete or adversarially influenced. Automated investigation techniques could also be misused. Accordingly, we position AIDA as a tool to support human analysts rather than enable autonomous decision-making, and we recommend access controls and human review for high-impact decisions. These safeguards limit foreseeable harms while preserving the defensive and reproducible value of the work.




%
\bibliographystyle{IEEEtran}
\bibliography{references}

\input{sections/appendix}

\end{document}

%% file: commands.tex
\usepackage{tikz}
\usepackage{amsmath}

\usepackage{filecontents}
\usepackage{tablefootnote}

\usepackage{microtype}
\usepackage{algorithm}
\usepackage[spaceRequire=false]{algpseudocodex}
\usepackage[utf8]{inputenc} 
\usepackage{listings}
\usepackage{xspace}
\usepackage{fvextra}
\usepackage{subcaption}
\usepackage{soul}
\usepackage{enumitem}
\usepackage{booktabs}
\usepackage{multirow}
\usepackage{tcolorbox}
\usepackage[subtle]{savetrees}
\usepackage{makecell}
\usepackage{colortbl}

\usepackage[capitalize,nameinlink]{cleveref}
\crefname{figure}{Fig.}{Figs.}
\crefname{listing}{Listing}{Listings}
\crefname{section}{\S}{Sections}
\crefname{table}{Tab.}{Tables}
\crefname{algorithm}{Algorithm}{Algorithms}
\crefname{line}{Line}{Lines}
\crefname{equation}{Eq.}{Eqs.}
\crefname{appendix}{Appendix}{Appendices}

\newcounter{rq}
\renewcommand{\therq}{\arabic{rq}}

\crefname{rq}{RQ}{RQs}
\Crefname{rq}{RQ}{RQs}
\crefformat{rq}{#2RQ#1#3}
\Crefformat{rq}{#2RQ#1#3}

\newenvironment{rqlist}{%
  \begin{list}{}{%
    \setlength{\leftmargin}{3.5em}%
    \setlength{\labelwidth}{3.0em}%
    \setlength{\labelsep}{0.5em}%
    \setlength{\itemsep}{1pt}%
    \setlength{\topsep}{2pt}%
    \setlength{\parsep}{0pt}%
    \setlength{\partopsep}{0pt}%
  }%
}{%
  \end{list}%
}

\newcommand{\rqitem}[2]{%
  \refstepcounter{rq}%
  \label{#1}%
  \item[\textbf{RQ\therq.}]%
  #2%
}

\newcommand{\myparagraph}[1]{\smallskip\noindent\textbf{#1.}}

\newcommand{\tool}{\textsc{Aida}\xspace}
\newcommand{\benchmark}{\textsc{Alert-Bench}\xspace}

\newif\ifshowcomment
\showcommenttrue

\ifshowcomment
    \newcommand{\pgao}[1] {{\footnotesize\color{red}[Peng: #1]}}
    \newcommand{\saimon}[1] {{\footnotesize\color{orange}[Saimon: #1]}}
\else
    \newcommand{\pgao}[1]{}
    \newcommand{\saimon}[1]{}
\fi

\newcommand{\eat}[1]{}

\definecolor{thesisblue}{RGB}{25,90,170}
\definecolor{adversaryred}{RGB}{190,60,30}       
\definecolor{judgegreen}{RGB}{30,130,60}
\definecolor{dpgray}{RGB}{90,90,110}

\definecolor{studybg}{HTML}{FCEDEC}
\definecolor{studyfg}{HTML}{9B3A32}
\definecolor{aidabg}{HTML}{EAF5EC}
\definecolor{aidafg}{HTML}{2F6B3B}
\definecolor{headergray}{HTML}{F3F4F6}

\definecolor{findingaccent}{RGB}{126,68,105}
\definecolor{findingbg}{RGB}{250,246,249}

\newcounter{finding}
\renewcommand{\thefinding}{\arabic{finding}}

\crefname{finding}{Finding}{Findings}
\Crefname{finding}{Finding}{Findings}

\newtcolorbox{finding}[1]{%
  enhanced,
  breakable,
  colback=findingbg,
  colframe=findingaccent,
  boxrule=0.8pt,
  arc=3pt,
  outer arc=3pt,
  left=6pt,
  right=6pt,
  top=5pt,
  bottom=5pt,
  boxsep=0pt,
  fontupper=\small,
  before skip=7pt,
  after skip=7pt,
  before upper={%
    \refstepcounter{finding}%
    \label{#1}%
    \setlength{\parskip}{3pt}%
    \setlength{\parindent}{0pt}%
    \textbf{Finding~\thefinding:}\hspace{0.35em}%
  }%
}

\usepackage[T1]{fontenc}
\usepackage{xcolor}
\usepackage{fvextra}
\usepackage{tcolorbox}
\usepackage[scaled=1.00]{inconsolata}

\tcbuselibrary{breakable,skins}

\definecolor{thesisblue}{RGB}{25,90,170}
\definecolor{adversaryred}{RGB}{190,60,30}
\definecolor{judgegreen}{RGB}{30,130,60}
\definecolor{dpgray}{RGB}{90,90,110}

\DefineVerbatimEnvironment{prompttext}{Verbatim}{
  fontsize=\small,
  formatcom=\ttfamily,
  baselinestretch=1.15,
  breaklines=true,
  breakanywhere=false,
  breakindent=0pt,
  breakautoindent=false,
  breaksymbolleft={},
  breaksymbolright={},
  xleftmargin=0pt,
  xrightmargin=0pt
}

\tcbset{
  agentprompt/.style={
    enhanced jigsaw,
    breakable,
    coltitle=white,
    fonttitle=\bfseries\small,
    boxrule=0.6pt,
    arc=2pt,
    left=6pt,
    right=6pt,
    top=7pt,
    bottom=7pt,
    toptitle=2pt,
    bottomtitle=2pt,
    before skip=10pt,
    after skip=10pt,
    attach boxed title to top left={
      yshift=-2mm,
      xshift=4mm
    }
  }
}

\newtcolorbox{strategistpromptbox}[1][]{%
  agentprompt,
  colback=thesisblue!7!white,
  colframe=thesisblue,
  title={#1},
  boxed title style={
    colback=thesisblue,
    colframe=thesisblue,
    arc=1.5pt,
    boxrule=0pt,
    left=5pt,
    right=5pt
  }
}

\newtcolorbox{adversarypromptbox}[1][]{%
  agentprompt,
  colback=adversaryred!7!white,
  colframe=adversaryred,
  title={#1},
  boxed title style={
    colback=adversaryred,
    colframe=adversaryred,
    arc=1.5pt,
    boxrule=0pt,
    left=5pt,
    right=5pt
  }
}

\newtcolorbox{judgepromptbox}[1][]{%
  agentprompt,
  colback=judgegreen!7!white,
  colframe=judgegreen,
  title={#1},
  boxed title style={
    colback=judgegreen,
    colframe=judgegreen,
    arc=1.5pt,
    boxrule=0pt,
    left=5pt,
    right=5pt
  }
}

%% file: sections/abstract.tex
Security operations centers (SOCs) must triage large volumes of alerts, most of which are benign, while missed attacks can remain uninvestigated. Tool-using large language model (LLM) agents can retrieve evidence during triage, but it remains unclear how reasoning strategies determine what to gather and when an investigation is sufficient to close an alert. We study five representative approaches spanning single-pass tool use, iterative retrieval, sampled investigations, self-review, and explicit verification. To support this study, we build \benchmark, an interactive benchmark that replays enterprise telemetry through a live SIEM and requires each system to retrieve evidence. Across 1,247 alerts from a four-day multi-stage attack scenario, every approach missed at least 40.4\% of attack-related alerts. Trace analysis shows that attack alerts are more likely to be dismissed when searches return no records, same-context review has negative net correction, and dismissal receives no consistently stronger investigation than escalation.

Based on these findings, we further design \tool (Adversarial Investigation and Dialectical Analysis), a multi-agent framework that requires an explicit proposed decision before independent challenge and stronger evidentiary requirements before dismissal. AIDA preserves investigation history in an append-only Investigation Ledger and keeps the challenge in a separate reasoning context. A separate Judge adjudicates the proposed decision and challenge against evidence, resolving the alert or requesting another round when evidence is missing. On the same alerts, AIDA achieves an F1 score of 0.958, compared with 0.371–0.744 for the studied approaches, and reduces the false-negative rate from 40.4\% to 3.1\% while escalating 18.4\% of alerts to analysts. These results show that structuring evidence retrieval and decision review can substantially improve agentic SOC triage.

%% file: sections/introduction.tex
\section{Introduction}
\label{sec:introduction}

Security operations centers (SOCs) receive large volumes of alerts from heterogeneous detection systems, yet the vast majority do not correspond to actionable threats~\cite{socalerts,harmony}. This volume is driven in part by detection rules (e.g., written in the Sigma format~\cite{sigma}) tuned for broad coverage, which repeatedly fire on common patterns and generate redundant alerts~\cite{ruling_the_unruly}. Many of these alerts are \emph{benign triggers}, where a rule correctly matches a suspicious pattern even though the underlying activity is legitimate in that environment and can be dismissed only after checking local context. At operational scale, security analysts process tens of thousands of alerts per day, while only a negligible fraction correspond to confirmed compromises~\cite{true_attacks_benign}. Most of that volume is absorbed by tier-1 triage, the first stage of the workflow, where each alert is either closed or escalated for deeper (tier-2 and tier-3) analysis~\cite{give_me_structure}. A closed alert may receive no further review, so dismissing a true attack allows it to proceed undetected, while escalating benign activity consumes analyst attention and further amplifies overload. Effective tier-1 triage must therefore minimize missed attacks and unnecessary escalations simultaneously, a requirement that existing automated approaches have not reliably achieved~\cite{tariq:fatigue}.

Prior automation reduces alert volume, but its analysis follows a predefined procedure that does not adapt the investigation to what remains unresolved. Log-based anomaly detection~\cite{deeplog,loganomaly} scores observed activity using learned detection models, contextual event correlation~\cite{deepcase,holmes} relates events across their surrounding activity, and provenance-based analysis~\cite{distdet,rapsheet,nodoze,flash,kairos} reasons over causal relationships captured in provenance graphs. These systems may incorporate contextual evidence, but the analysis pipeline determines how that evidence is processed and does not decide what additional information to gather when the available evidence is insufficient. Tool-using agents built around large language models (LLMs) have the potential to support this iterative investigation~\cite{react,toolformer}.
Security research has applied LLM-based systems in other domains, such as penetration testing~\cite{pentestgpt}, threat intelligence~\cite{ctinexus}, insider threat detection~\cite{auditllm}, and incident response~\cite{llms_in_incident_response,llms_in_the_soc_study}. However, how LLM-based agents can support efficient and reliable tier-1 triage remains largely underexplored. Tool access alone does not ensure reliable investigation, as prior work shows that LLM judgments are sensitive to surrounding context~\cite{prosa,scalar_prompt,lostmiddle} and that self-checking often fails to correct unsupported conclusions~\cite{closer_self_verif,llms_cant_selfcorrect,llms_cant_reason_errors}.

Whether an LLM agent can triage alerts therefore depends on how its reasoning is organized. To understand the capabilities and limitations of LLM-based agents for tier-1 triage, we take a first step toward a comprehensive study of five representative approaches spanning widely used reasoning paradigms in tool-using agents. Direct-Prompt reads the alert, makes a small number of tool calls, and returns a decision in one pass~\cite{gpt3}. ReAct adds iterative tool use so that each result guides the next action~\cite{react}. Self-Consistency runs several independent investigations and selects the majority decision~\cite{selfconsistency}. Self-Refine reviews and revises its own decision~\cite{selfrefine}. Chain-of-Verification (CoVe) generates and answers checking questions before revising the decision~\cite{cove}. Together, these approaches cover single-pass tool use, iterative retrieval, sampled investigations, self-review, and explicit verification. We therefore ask: To what extent can these approaches triage alerts reliably?

To faithfully evaluate these approaches, we need an environment that reproduces tier-1 investigation, where a system starts from one alert and decides what surrounding evidence to inspect before dismissing or escalating it. Existing benchmarks evaluate related security tasks. CyberSOCEval focuses on malware analysis and threat-intelligence reasoning~\cite{cybersoceval}. ExCyTIn-Bench evaluates agents that retrieve security logs to answer investigation questions~\cite{excytin}, whereas our task is to decide whether a detector-generated alert warrants escalation. SIABench evaluates alert triage using tasks that pair an alert with a selected PCAP artifact for analysis~\cite{siabench}. The system can inspect that artifact with security tools, but it is not evaluated on deciding what evidence to retrieve from the broader telemetry surrounding the alert. To support this evaluation, we build \benchmark, an interactive alert-triage benchmark in which each system starts from one detector-generated alert and searches a live Security Information and Event Management (SIEM) deployment for supporting evidence. AIT-LDSv2~\cite{aitldsv2} provides recorded host and network telemetry together with event-level attack annotations. We replay this telemetry through the SIEM to generate alerts from both attack and routine activity, then use the alert-to-event matching procedure from AIT-ADS~\cite{aitads} to assign ground-truth labels, which remain hidden during investigation. These source datasets provide the recorded telemetry and attack annotations, while \benchmark turns them into an evaluation environment where the system investigates each alert through the SIEM before reaching a triage decision, mirroring the evidence-gathering process of a security analyst.

Our study shows that these approaches can carry out multi-step alert investigation, while revealing recurring limitations in how they gather and use evidence. The investigation traces show three recurring behaviors. (1) Attack-related alerts are much more likely to be dismissed when the agents' chosen searches for surrounding activity return no records. Four of the five approaches dismiss at least 88\% of attack alerts in such investigations. An empty result only shows that a particular search found nothing, yet the agents often treat it as support for closing the alert without establishing that enough relevant evidence has been examined. (2) Same-context review often makes the decision worse. Self-Refine corrects 16 incorrect decisions but changes 79 correct ones into errors, while CoVe corrects 25 and breaks 132. These review stages often reinterpret evidence already present in the reasoning context instead of correcting the decision with new evidence. (3) Dismissed alerts do not receive a stronger investigation even though dismissal ends the alert's triage path. Before dismissing or escalating an alert, the approaches perform roughly the same number of searches. Yet four of the five retrieve fewer records before dismissal, meaning that alerts are often closed after less evidence has actually been gathered. \emph{These findings show that the organization of evidence retrieval and decision review strongly shapes whether additional reasoning corrects an error or reinforces it.}

We take one step further and use these findings to design \tool (Adversarial Investigation and Dialectical Analysis), a multi-agent architecture for alert triage that structures evidence gathering, independent challenge, and separate adjudication. To enforce this structure, we introduce the \emph{Dialectical Commitment Protocol}, which coordinates four agents, the Strategist, Executor, Adversary, and Judge, through an append-only \emph{Investigation Ledger} (Figure~\ref{fig:architecture}). The ledger preserves retrieved evidence and records how proposed decisions are challenged across investigation rounds. The Strategist identifies the evidence needed to assess the alert and sends requests to the Executor, which performs the searches, handles failures, and writes normalized results to the ledger. After gathering sufficient evidence or reaching the per-round tool budget, the Strategist records a \emph{thesis}, an explicit proposed decision supported by the retrieved evidence. Committing the thesis before challenge gives the Adversary a concrete claim to evaluate from a separate reasoning context, addressing the negative effects of same-context review observed in our study. The Adversary records a \emph{challenge} when the evidence contradicts the thesis or when information needed to resolve the alert is still missing. The Judge then evaluates the thesis and challenge against the collected evidence. Because dismissal ends the alert's triage path, the Judge requires stronger evidence before closing an alert. When important evidence is still missing, it requests another investigation round and identifies what should be gathered next. The next round continues from the existing ledger, preserving the investigation history while allowing the proposed decision to change as new evidence is retrieved.

To evaluate \tool's performance and how well it addresses the limitations identified in our study, we compare it with the five studied approaches using the same underlying model and evidence interface. \tool achieves an F1 score of 0.958 with 0.969 recall, compared with F1 scores of 0.371--0.744 for the studied approaches. Their false-negative rates range from 40.4--70.2\%, while \tool reduces this rate to 3.1\%. 
This higher attack coverage increases the fraction of alerts sent to analysts to 18.4\%, compared with 8.5--14.0\% for the studied approaches.
We also evaluate the contribution of separate adjudication. It corrects 31 incorrect proposed decisions without reversing a correct one, while removing the Judge agent in the ablation reduces F1 from 0.952 to 0.895. Finally, we evaluate whether strong performance extends beyond the reference model. The open-weight Qwen3.5 9B model reaches an F1 score of 0.952 with \tool.

\myparagraph{Contributions}
This paper makes the following contributions.

\begin{enumerate}[align=left,labelsep=0pt,labelwidth=\parindent,leftmargin=\parindent]

  \item \textbf{\benchmark, an interactive alert-triage benchmark.}
  We build a construction pipeline and evaluation harness that turn recorded enterprise telemetry into detector-generated alerts with hidden ground truth that systems investigate through a live SIEM. Unlike existing benchmarks that pair an alert with selected evidence for analysis, \benchmark begins from the alert alone and requires each system to decide what supporting evidence to retrieve before reaching a triage decision~\cite{siabench}. This enables controlled evaluation of both evidence gathering and decision making during alert investigation.

  \item \textbf{A systematic study of LLM reasoning approaches for interactive alert triage.}
  We conduct a systematic comparative study of representative tool-using LLM reasoning approaches for interactive SOC alert triage. We evaluate five approaches under identical alerts, model choice, evidence access, and tool capabilities. By analyzing both their decisions and investigation traces, we identify where these approaches fail during evidence gathering and review. Attack-related alerts are much more likely to be dismissed when the agents' searches return no records, while same-context review often introduces more errors than it corrects. We also find that dismissal does not receive stronger investigation than escalation, with four approaches retrieving fewer records before closing an alert.

  \item \textbf{\tool, a multi-agent architecture for structured alert investigation.}
  Building on the insights from our study, we design \tool, a multi-agent architecture that organizes alert triage around separate investigation and independent challenge. To realize this design, we introduce the \emph{Dialectical Commitment Protocol}, which requires an explicit proposed decision before challenge and assigns final decision authority to a separate Judge. An append-only Investigation Ledger preserves the investigation across rounds, allowing the Judge to request additional evidence when important information is still missing.

\end{enumerate}

%% file: sections/background.tex
\section{Background and Related Work}
\label{sec:background}

\subsection{Alert Triage and Its Automation}

We consider enterprise environments in which alerts arise from multi-stage attack scenarios spanning reconnaissance through privilege escalation. Triage is the task of identifying malicious activity within predominantly benign telemetry, and resolving a single alert requires correlating activity across hosts and over time~\cite{socalerts,true_attacks_benign,give_me_structure}. Studies of analyst performance under these conditions report both missed attacks and escalation fatigue as alert volume rises and context stays incomplete~\cite{sundaramurthy:burnout,layman:falsepositive,tariq:fatigue}.

Automated approaches to this workload fall into three families. Log anomaly detection models system logs as event sequences and flags departures from learned behavior~\cite{deeplog,loganomaly,logbert}. Contextual correlation groups related events, with DeepCASE clustering correlated security-event sequences for operator review and HOLMES relating alerts to attack stages~\cite{deepcase,holmes}. Provenance-based systems reason over causal structure~\cite{distdet,kairos,nodlink}. NoDoze scores alerts against historical context over dependency graphs, RapSheet builds tactical provenance graphs over EDR alerts, and FLASH applies graph representation learning to provenance for intrusion detection~\cite{nodoze,rapsheet,flash}. Their effectiveness also depends on the coverage of the learned representation or historical context, which can degrade under novel attack behavior and benign activity that resembles an attack~\cite{arp:dosdonts}. Their decisions are based on representations constructed by a fixed analysis procedure. Once that representation is formed, the system does not decide what additional evidence to retrieve in response to uncertainty.

\subsection{LLM Agents in Security Operations}

Large language models support security automation in which a system retrieves and analyzes evidence during inference. Beyond producing a label, this lets a triage system state an evidence-grounded rationale in natural language that an analyst can inspect and override, which is the property that motivates our use of language models. LLM-based systems have been applied to penetration testing~\cite{pentestgpt}, threat intelligence~\cite{ctinexus}, insider threat detection~\cite{auditllm}, and incident response~\cite{llms_in_incident_response,llms_in_the_soc_study}. In SOC workflows, such systems can query logs and retrieve related context before reaching a decision, which turns triage from static classification into iterative investigation.

Multi-agent architectures extend LLM-based investigation by assigning different agents to evidence collection and reasoning~\cite{du:debate,mad,autogen}. In alert triage, this allows coordination across several steps of evidence collection and analysis over logs and alerts. Most such systems coordinate through a shared, continuously updated context, where each agent reads and writes the same state, so the material an agent works from already contains the steps taken before it. Studies of multi-agent LLM systems report that agents conform to prior responses~\cite{weng2025conformity,consensagent} and can shift toward incorrect conclusions during interaction~\cite{wynn2025talk}, and that an error introduced early propagates through the shared state~\cite{cemri2025multi}. These reports concern general reasoning and coordination tasks. Section~\ref{sec:systematic_study} measures the effect of this shared reasoning context on alert triage, and Section~\ref{sec:system} presents the structure we build in response.

\subsection{Evaluation and Benchmarking}

CyberSOCEval evaluates malware analysis and threat-intelligence reasoning~\cite{cybersoceval}. Cybench evaluates cybersecurity agents on interactive capture-the-flag challenges~\cite{cybench}, where success means solving the challenge and recovering a flag. \benchmark evaluates defensive alert-triage decisions over enterprise telemetry.
ExCyTIn-Bench~\cite{excytin} evaluates multi-hop threat investigation using graph-derived questions that agents answer by querying security-log tables. It shares our emphasis on retrieving evidence, but its target is the answer to an investigation question. \benchmark instead evaluates whether a detector-generated alert should be dismissed or escalated, across both attack-related and benign activity.
SIABench generates detector alerts from selected PCAP segments and supplies the associated artifact for investigation~\cite{siabench}. Its agents can inspect that artifact with security tools. \benchmark requires systems to retrieve supporting evidence from the surrounding telemetry through the SIEM interface, as described in Section~\ref{sec:benchmark}.

%% file: sections/benchmark.tex
\section{\benchmark}
\label{sec:benchmark}

\begin{figure}[t]
\centering
\includegraphics[width=\columnwidth]{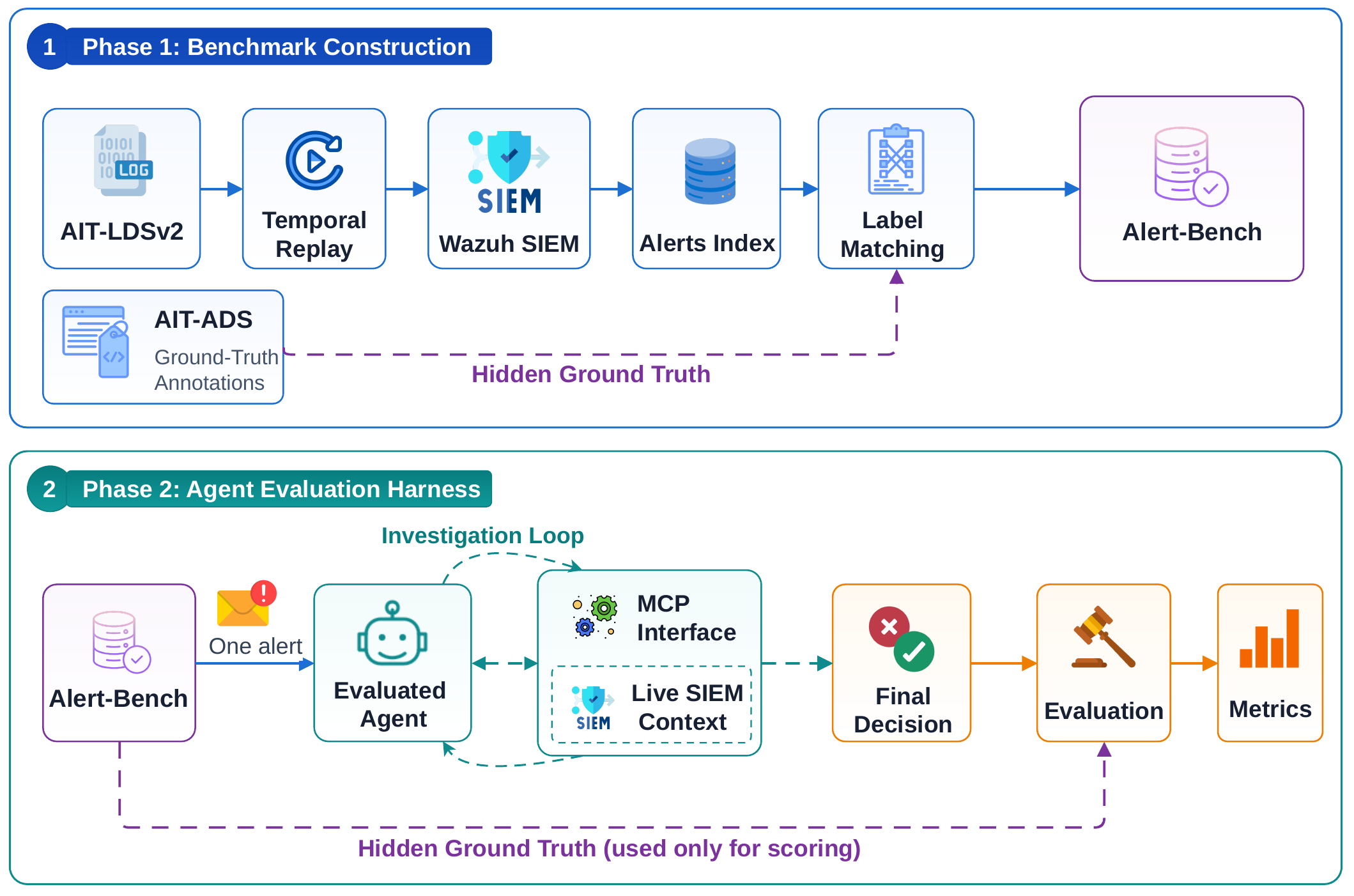}
\caption{Construction and evaluation pipeline of \benchmark. Phase 1 replays AIT-LDSv2 telemetry through a Wazuh SIEM and aligns the resulting alerts with the event-based ground truth used by AIT-ADS. Phase 2 gives each evaluated system one alert and access to the replayed telemetry through a fixed MCP interface. Ground-truth labels remain hidden and are used only for scoring.}
\label{fig:alertbench_pipeline}
\end{figure}

\benchmark turns the AIT datasets into an executable alert-triage benchmark. AIT-LDSv2 provides recorded enterprise telemetry and event-level attack annotations from multi-stage attack scenarios~\cite{aitldsv2,aitldsv2_dataset}, while AIT-ADS provides the procedure for aligning detector alerts with these annotations~\cite{aitads}. These resources provide the telemetry and ground truth, but not an interactive task in which a system must determine what evidence to retrieve before resolving an alert. \benchmark adds this evaluation environment. Each instance begins with one detector-generated alert and no preselected evidence bundle. The system must retrieve any additional context from the replayed telemetry through a fixed SIEM interface, while the associated ground-truth label remains hidden until scoring.

\subsection{Environment Construction and Ground Truth}
\label{subsec:bench_construction}

\myparagraph{Replay environment}
Figure~\ref{fig:alertbench_pipeline} shows the construction process. We replay AIT-LDSv2 log data in timestamp order into a Wazuh~\cite{wazuh} deployment. The source data also contain Suricata~\cite{suricata} alerts, which are ingested with the host telemetry. Wazuh processes the incoming logs through its rule engine~\cite{wazuh_ruleset} and indexes the resulting alerts. Original timestamps and host identities are preserved, leaving the surrounding activity searchable across hosts and over time during investigation.

\myparagraph{Ground-truth labeling}
We assign ground truth using the event-based labeling procedure published by the AIT-ADS authors~\cite{aitads,aitads_code}. Wazuh alerts are matched to the corresponding annotated events in AIT-LDSv2 using their procedure. Suricata alerts use the authors' flow-based matching, while audit alerts spanning several records use the corresponding event-matching procedure. If a matched event carries multiple attack labels, a fixed priority order maps it to one attack phase for phase-level analysis.

We retain alerts associated with the replayed telemetry and exclude alerts generated by unrelated activity on the SIEM host. This filtering uses alert source and host information, not the ground-truth class. Alerts matching an attack annotation are labeled attack-related; unmatched alerts are retained as benign under the AIT-ADS event-based labeling convention~\cite{aitads}. Labels are stored outside the investigation interface and used only for scoring.

\subsection{Evaluation Corpus}
\label{subsec:bench_corpus}

\myparagraph{Scenario selection}
The construction pipeline applies to all eight AIT-LDSv2 scenarios, whose alert distributions differ with campaign activity and background traffic~\cite{aitldsv2,aitads}. For the comparative study, we use \textit{russellmitchell}. Its detector output is benign-dominated and has the smallest published alert volume among the eight AIT-ADS scenarios, making repeated evaluation across several agent designs tractable. This selection uses only properties of the source scenario and does not depend on results from any evaluated method.

The replay covers four days and contains a multi-stage campaign spanning directory enumeration and WordPress vulnerability scanning, with additional service-scanning and privilege-escalation activity. We preserve the scenario's class distribution instead of constructing a balanced workload. The same construction and labeling procedure applies to the remaining AIT-LDSv2 scenarios without changing the benchmark task.

\myparagraph{Sampling and composition}
The comparative evaluation uses a fixed sample of 1{,}247 alerts from \textit{russellmitchell}. We use stratified sampling to remain close to the source class distribution while retaining its attack phases. Attack-related alerts are stratified by attack phase and benign alerts by detection rule. Sampling is deterministic under a fixed random seed. Every evaluated method receives the same sample, and the smaller ablation and model subsets are drawn from it to preserve matched comparisons.

The resulting corpus contains 225 attack-related and 1{,}022 benign alerts. Among the attack alerts, 126 correspond to directory enumeration, 94 to WordPress vulnerability scanning, 3 to service scanning, and 2 to privilege escalation through sudo. Table~\ref{tab:dataset} gives the distribution by detection rule. Attack alerts concentrate in two web-server rules triggered by HTTP 400 responses during reconnaissance, while benign alerts are dominated by successful Dovecot logins, generic IDS events, and Suricata TLS alerts.

\begin{table}[t]
\centering
\small
\caption{Composition of the 1{,}247-alert evaluation sample by detection rule. Attack alerts concentrate in web-server reconnaissance rules, while benign alerts span authentication and monitoring activity.}
\label{tab:dataset}
\setlength{\tabcolsep}{4.5pt}
\begin{tabular}{llrr}
\toprule
\textbf{Rule} & \textbf{Description} & \textbf{Attack} & \textbf{Benign} \\
\midrule
9701  & Dovecot login success        & 0   & 433 \\
20101 & Generic IDS event            & 0   & 275 \\
86601 & Suricata TLS handshake       & 3   & 273 \\
31101 & Web server 400 error         & 187 & 0   \\
31151 & Repeated 400s, same source   & 33  & 0   \\
52507 & ClamAV database update       & 0   & 25  \\
\midrule
\multicolumn{2}{l}{Other (14 rules)}  & 2   & 16 \\
\midrule
\multicolumn{2}{l}{\textbf{Total}}   & \textbf{225} & \textbf{1{,}022} \\
\bottomrule
\end{tabular}
\end{table}

\subsection{Interactive Task and Harness}
\label{subsec:bench_harness}

\myparagraph{Investigation interface}
The unit of evaluation is one alert together with any surrounding telemetry the system chooses to retrieve. Each system receives the alert record and investigates it through a fixed interface built on the Model Context Protocol (MCP)~\cite{mcp}. The alert timestamp anchors retrieval windows. The interface exposes four investigation operations and a host-inventory lookup.

\begin{itemize}[align=left,labelsep=4pt,leftmargin=*,itemsep=1pt,topsep=2pt]
  \item \texttt{get\_wazuh\_alerts} retrieves alerts by rule identifier, host, and time window to examine related alert activity.
  \item \texttt{search\_security\_events} searches security records for an entity such as an IP address or user name to follow it into related activity.
  \item \texttt{get\_wazuh\_alert\_summary} aggregates alert counts over a selected field and time window to compare activity frequency.
  \item \texttt{analyze\_alert\_patterns} identifies spikes and rare entity combinations within a selected window for broader investigation.
\end{itemize}

A separate host-inventory operation maps SIEM agent identifiers to their hosts. All outputs are restricted to the replayed corpus, and no operation exposes ground-truth labels. Every system therefore begins from the same alert with the same evidence access and tool capabilities, while deciding independently what to retrieve and when the investigation is sufficient.

Each investigation ends with a decision to escalate or dismiss the alert together with a justification. The harness records the decision and scores it against the hidden ground truth after the investigation ends.

\myparagraph{Investigation traces}
The harness records each investigation as an ordered transcript and separately counts its tool calls. Each transcript preserves the queries and arguments, returned results, and intermediate or final decisions. Rejected queries retain the server error, while successful queries with no matches are recorded as empty results. This distinction separates execution failures from successful searches that retrieve no records and supports the trace analysis in Section~\ref{subsec:failure_analysis}.

\subsection{Scoring}
\label{subsec:bench_scoring}

\myparagraph{Classification metrics}
We treat attack-related alerts as the positive class. We report precision, recall, and F1, together with accuracy and confusion-matrix counts. Recall measures the fraction of attack alerts that reach an analyst, while precision measures how often an escalated alert is attack-related.

\myparagraph{Operational metrics}
We report the false-negative rate (FNR),
$\text{FN}/(\text{TP}+\text{FN})$, as the fraction of attack alerts incorrectly dismissed, and Investigation Load (IL),
$(\text{TP}+\text{FP})/\text{Total}$, as the fraction of alerts sent to an analyst. These metrics must be interpreted together. A system that dismisses every alert would have zero Investigation Load while missing every attack.

We also report negative predictive value (NPV),
$\text{TN}/(\text{TN}+\text{FN})$, which measures how often a dismissed alert is actually benign. NPV therefore indicates how reliable the system's closure decisions are.

\myparagraph{Cost metrics}
For each alert, the harness records token use, tool-call count, and end-to-end wall-clock latency. We report these separately from triage quality so that computational effort can be compared with decision performance.

%% file: sections/systematic_study.tex
\section{Systematic Study}
\label{sec:systematic_study}

\benchmark lets us study representative reasoning approaches in an interactive alert-triage setting. We first measure the triage quality they achieve and then analyze their investigation traces to understand how they gather and use evidence as an investigation progresses. This section reports both parts of the study.

\subsection{Study Design}
\label{subsec:study_design}

\myparagraph{Research questions}
The study asks the following research questions.

\begin{rqlist}
\rqitem{rq:study-triage}{
  To what extent can representative reasoning approaches reliably triage SOC alerts through interactive investigation?
}

\rqitem{rq:study-evidence}{
  How do these approaches gather and use evidence as an alert investigation progresses?
}
\end{rqlist}

\myparagraph{Approaches}
We evaluate the five reasoning approaches introduced in Section~\ref{sec:introduction}. They span single-pass reasoning, iterative tool use, sampled investigations, self-critique, and explicit self-verification. Their structures and reasoning patterns are summarized in Table~\ref{tab:baselines} in Appendix~\ref{app:reasoning-approaches}. All five start from the same alerts and use the same model through the same investigation interface.

We treat each approach as a complete reasoning strategy. Their control flow and computational budgets differ because those choices are part of the strategy itself. We therefore use the trace analysis in Section~\ref{subsec:failure_analysis} to identify behaviors that recur within and across approaches instead of attributing aggregate differences to one component alone.

\myparagraph{Configuration}
Tool budgets are set high enough for each approach to complete its prescribed interaction pattern. Direct-Prompt may issue three calls and ReAct may issue five. Self-Refine receives seven calls, while CoVe receives ten. Self-Consistency runs three ReAct trajectories with five calls each and resolves the final decision by majority vote. Average tool use remains below the corresponding cap for every approach.

All approaches receive the same tool definitions and parameter descriptions. Decoding uses a temperature of 0.2 except for Self-Consistency, which uses 0.7 because its mechanism depends on diversity across sampled trajectories.

\myparagraph{Experimental setup}
All experiments run on a dedicated Ubuntu 20.04 server with two AMD EPYC 7313 processors, 1.0\,TiB of RAM, and one NVIDIA A100 80\,GB GPU. The server hosts the Wazuh SIEM and the MCP investigation service together with the evaluation harness. GPT-4.1 (gpt-4.1-2025-04-14)~\cite{gpt41_model} is the reference model for this study. Every approach is evaluated on the same 1{,}247 alerts through the interface defined in Section~\ref{subsec:bench_harness}.

The study contains 6{,}235 investigations, one run of each approach on every alert in the evaluation sample. This matched design supports both aggregate comparison and per-alert analysis without differences in the underlying workload.

\myparagraph{Statistical procedure}
We report 95\% Wilson score confidence intervals for rate metrics~\cite{wilson1927}. Because every approach is evaluated on the same alerts, comparisons are paired at the alert level. We use McNemar's exact test for differences in per-alert correctness. Differences in F1 are evaluated with a stratified paired bootstrap that preserves the positive and negative counts. We use Spearman's rank correlation for ordered retrieval counts and the Mann--Whitney $U$ test for retrieval-volume comparisons. All procedures use a fixed random seed.

\myparagraph{Scope}
The study focuses on reasoning strategies that can operate directly over the benchmark interface without task-specific training. Provenance-based methods require causal graphs that are not available in the source telemetry. Supervised classifiers require labeled training data from the deployment environment, which changes the zero-training investigation setting studied here.

\subsection{Aggregate Results}
\label{subsec:aggregate}

\begin{figure*}[t]
\centering
\includegraphics[width=\textwidth]{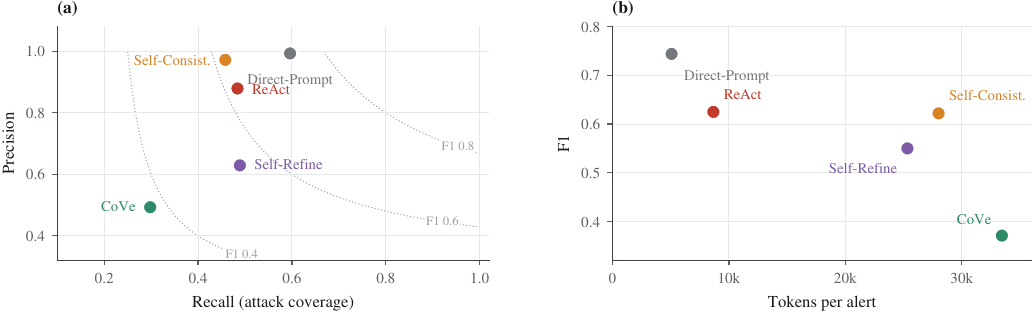}
\caption{Aggregate results of the five studied approaches. Panel (a) shows their precision--recall operating points with F1 contours. Panel (b) compares triage quality against computational cost using tokens per alert. Changing the reasoning structure moves the operating point, but higher computational cost does not correspond to better triage quality.}
\label{fig:systematic_aggregate}
\end{figure*}

\myparagraph{\cref{rq:study-triage}. Triage effectiveness}
Figure~\ref{fig:systematic_aggregate}(a) shows that none of the five approaches reaches both high precision and high recall. Direct-Prompt obtains the highest F1 score at 0.744 with 0.596 recall, while the remaining approaches range from 0.371 to 0.625 F1. Even the strongest approach therefore dismisses 40.4\% of attack-related alerts.

The common failure is missed attacks. Self-Consistency retains high precision but reduces recall to 0.458, while Self-Refine and CoVe lose substantial precision without recovering the attack coverage achieved by Direct-Prompt. None of the additional reasoning structures improves on the single-pass result in this evaluation.

Direct-Prompt also differs significantly from each of the other four approaches in per-alert correctness under McNemar's exact test. The smallest difference is against Self-Consistency, with a $p$-value of $1.1\times10^{-3}$. We do not interpret the ordering among the remaining approaches where pairwise differences are unresolved at this sample size.

Investigation Load must be interpreted with the false-negative rate. Self-Consistency sends only 8.5\% of alerts to analysts, the lowest load in the study, but dismisses 54.2\% of attack-related alerts. A system that dismissed every alert would report zero Investigation Load while missing every attack. A lower Investigation Load therefore does not by itself indicate better triage.

\begin{finding}{find:triage-effectiveness}
Reasoning structure changes the precision--recall operating point, but none of the five approaches achieves high attack coverage. The best F1 score is 0.744, and every approach dismisses at least 40.4\% of attack-related alerts.
\end{finding}

\myparagraph{Computational cost}
Figure~\ref{fig:systematic_aggregate}(b) shows that greater computational effort does not correspond to better triage quality. Direct-Prompt uses 5{,}075 tokens per alert and achieves the highest F1 score. Self-Consistency uses 28{,}002 tokens yet misses 31 more attacks, while CoVe uses 33{,}438 tokens and achieves the lowest F1 score of 0.371. Several approaches also issue substantially more tool calls without improving recall.

These results do not establish that additional computation is harmful by itself. They show that spending more computation through sampling or self-review is not sufficient to improve triage.

\begin{finding}{find:cost-quality}
Greater computational cost does not correspond to better triage quality. Direct-Prompt uses the fewest tokens and achieves the highest F1 score, while CoVe uses the most tokens and achieves the lowest.
\end{finding}

\subsection{Trace Analysis}
\label{subsec:failure_analysis}

\myparagraph{\cref{rq:study-evidence}. Investigation behavior}
We analyze their traces to characterize how they gather evidence and how retrieved results shape later investigation steps and final decisions.

\myparagraph{Method}
Retrieval counts cover only the investigation stage, before any review or verification step, while computational-cost measurements count all tool calls issued by the complete approach. We parse all 6{,}235 transcripts with a structural parser keyed to the server's output markers. Decisions reconstructed from the parsed traces reproduce the aggregate confusion matrix for every approach exactly. We also re-read the transcripts with a second parser that shares no code with the first and recovers retrieval outcomes by decoding each response payload instead of matching its markers. The two parsers agree on the retrieval state of all 6{,}235 traces.


\begin{figure}[t]
\centering
\includegraphics[width=0.9\columnwidth]{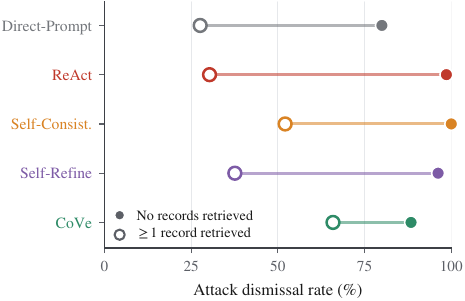}
\caption{Attack dismissal rate by retrieval state. When an investigation retrieves no records, the five approaches are substantially more likely to dismiss attack alerts than when at least one query returns records.}
\label{fig:retrieval_state}
\end{figure}

\myparagraph{Empty retrieval and attack dismissal}
Figure~\ref{fig:retrieval_state} shows that whether an investigation retrieves any records is strongly associated with its attack decision. ReAct dismisses 98.6\% of attacks when no records are retrieved, compared with 30.3\% when at least one query returns records. Self-Refine changes from 96.2\% to 37.6\%, while Direct-Prompt changes from 80.0\% to 27.6\%. Across the five approaches, 211 of the 602 false-negative investigations end without retrieving any records.

The benign alerts provide a control. Whether records are retrieved changes the benign dismissal rate by at most 2.8 percentage points. Retrieval state therefore has little effect when dismissal is already correct, while it is strongly associated with false negatives on attack alerts.

An empty query can provide valid negative evidence about the specific predicate that was searched. The problem appears when empty results support dismissal without a separate check that the investigation searched enough of the relevant context. None of the five approaches contains such a check. Class prevalence can make this behavior appear effective overall. No-record investigations produce 937 correct benign dismissals and 211 false-negative dismissals, concentrating the cost of this behavior on attack alerts.

\begin{figure}[!t]
\centering
\includegraphics[width=0.98\columnwidth]{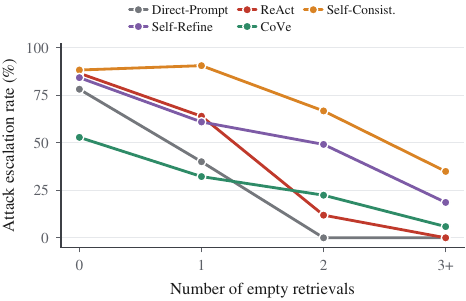}
\caption{Attack escalation as empty retrievals accumulate. Across most approaches, escalation falls as investigations encounter more empty results.}
\label{fig:empty_retrievals}
\end{figure}

\myparagraph{Effect of continued retrieval}
Figure~\ref{fig:empty_retrievals} shows that approaches performing more retrieval often continue into progressively less informative investigation states. For ReAct, escalation falls from 86.5\% when no query is empty to 11.9\% after two empty results. Self-Refine falls from 84.2\% to 49.1\% over the same range and reaches 18.6\% after three or more. Across all 1{,}125 attack investigations, the rank correlation between empty-result count and dismissal is 0.383 at $p<10^{-40}$.

Direct-Prompt and ReAct provide a useful comparison because they use the same model and investigation interface but differ in how long they continue retrieving. On attack alerts where at least one query returns records, Direct-Prompt escalates 72.4\% and ReAct escalates 69.7\%. On the 120 attack alerts where both retrieve records, they escalate 85 and 79 respectively. Their decisions are therefore similar once retrieval returns records. ReAct reaches the no-record state much more often, which helps explain why its additional retrieval does not improve the aggregate result.

\begin{finding}{find:empty-retrieval}
Empty retrievals are strongly associated with attack dismissal, and additional retrieval does not reliably correct this pattern. Four of the five approaches dismiss at least 88\% of attack alerts when no query retrieves a record, while dismissal increases with empty-result count across attack investigations.
\end{finding}


\myparagraph{Same-context self-review}
Self-Refine changes 95 decisions during its review stage. Only 16 changes repair an incorrect decision, while 79 turn a correct one into an error. CoVe changes 157 decisions, repairing 25 and breaking 132. Their F1 scores fall from 0.665 to 0.550 for Self-Refine and from 0.669 to 0.371 for CoVe after review.

The errors move in different directions. Self-Refine changes 73 dismissals into escalations, increasing its false positives from 8 to 65. CoVe changes 79 escalations into dismissals and reduces its true positives from 121 to 67. For Self-Refine, 74 of its 95 reversals occur without retrieving a new record. Its review therefore often changes the interpretation of information already present in the reasoning context. Despite moving errors in different directions, both review stages reduce the number of correct decisions.

\begin{finding}{find:self-review}
Same-context self-review has negative net correction. Self-Refine breaks 79 correct decisions while repairing 16, and CoVe breaks 132 while repairing 25.
\end{finding}


\myparagraph{Investigation effort before dismissal}
A dismissal ends the alert's triage path, while an escalation passes the alert to an analyst. The evidence supporting dismissal should therefore be at least as complete as the evidence supporting escalation. The studied approaches do not show this asymmetry. Across all five, the ratio between tool calls issued before dismissal and escalation ranges from 0.86 to 1.25.

The difference becomes clearer when we count only queries that return records. ReAct obtains 0.92 non-empty results before dismissal and 1.24 before escalation. Self-Consistency obtains 2.84 versus 3.66, while CoVe obtains 1.68 versus 2.55. Each difference is significant at $p<10^{-5}$. Self-Refine shows a smaller gap of 1.35 versus 1.61 at a $p$-value of 0.002. Only Direct-Prompt shows no difference in successful retrieval between the two decisions.

For four of the five approaches, the decision that ends analyst review therefore follows fewer successful retrievals. None of the five designs explicitly requires a more complete investigation before dismissal.

\begin{finding}{find:dismissal-effort}
The studied approaches do not consistently investigate more before dismissal than escalation. Across the five approaches, the number of tool calls issued before the two decisions differs by at most 25\%.
\end{finding}

\subsection{Error Overlap and Interpretation}
\label{subsec:error_concentration}

\myparagraph{Decision overlap across approaches}
The three findings under \cref{rq:study-evidence} characterize recurring investigation behaviors. We next examine whether the resulting attack decisions are shared across approaches or depend on how the investigation is organized.

Of the 225 attack alerts, 26 are missed by all five approaches and only 32 are caught by all five. Across their combined decisions, at least one approach escalates 199 attacks, corresponding to a recall of 0.884. The best individual approach reaches only 0.596 recall. Most attacks missed by one approach are therefore escalated by another under the same evidence interface. Agreement is much stronger on benign alerts, where all five dismiss 879 of the 1{,}022 alerts and no alert is escalated by all five. Much of the attack-recall loss therefore depends on how the investigation is organized.

\myparagraph{Coverage by attack phase}
The two dominant attack phases show the same limitation. Directory enumeration contributes 126 attack alerts and WordPress vulnerability scanning contributes 94, with best recall of 0.683 and 0.500 respectively. Twenty of the 26 attacks missed by every approach come from WordPress scanning, while four come from directory enumeration. Service scanning and privilege escalation contain only three and two alerts, so we do not draw comparative conclusions from them.

\myparagraph{Exact duplicate ambiguity}
We also test whether shared failures can be explained by attack alerts whose local content is identical to benign activity. For each attack alert, we mask the syslog timestamp and process identifier and search the replay corpus for an otherwise identical log line carrying the opposite label. Three of the 225 attack alerts meet this condition, including two of the 26 missed by every approach.

%% file: sections/system_design.tex
\section{Design of \tool}
\label{sec:system}

\subsection{Design Rationale}
\label{subsec:rationale}

Our study shows that the studied approaches do not provide reliable triage, and additional computation alone does not resolve this limitation. The trace analysis further shows that incomplete retrieval can support attack dismissal without establishing sufficient context, while same-context review can reinforce or introduce errors. Dismissal also receives no consistently stronger investigation than escalation, even though an incorrect dismissal may close a true attack without further review. We use these findings to derive the design of \tool (Adversarial Investigation and Dialectical Analysis), a multi-agent alert-triage framework that structures investigation around explicit commitment, independent challenge, and separate adjudication. \tool implements these ideas through three protocol properties that directly follow from the study findings. 

\myparagraph{Explicit Thesis Commitment}
Independent review requires a concrete conclusion to evaluate. \tool therefore requires the Strategist to commit a \emph{thesis}, a proposed classification supported by cited evidence, before challenge begins. The Strategist may commit once it considers the collected evidence sufficient. If it reaches the per-round tool budget first, the Forced Commitment Gate ends retrieval for that round, and the next Strategist turn produces a thesis from the evidence already collected. Every round therefore produces a fixed claim that can be evaluated from a separate reasoning context before the investigation proceeds.

\myparagraph{Independent Challenge and Adjudication}
\Cref{find:self-review} shows that same-context self-review produces negative net correction. \tool therefore separates the production and evaluation of the proposed classification. The Adversary receives the committed thesis and shared evidence without the Strategist's private reasoning, and records a challenge to the thesis. The Judge then evaluates the thesis and challenge against the evidence and makes the final decision or requests another investigation round. This structure gives the proposed classification an independent review and gives final decision authority to a separate role, a separation also evaluated in evidence-based debate~\cite{debate_truthfulness}.

\myparagraph{Asymmetric Evidentiary Thresholds}
\Cref{find:empty-retrieval,find:dismissal-effort} show that attack dismissal is closely associated with empty retrievals, while dismissal receives no consistently stronger investigation than escalation. \tool therefore applies different evidentiary requirements to the two outcomes. Escalation requires concrete evidence supporting malicious activity. Dismissal requires affirmative benign evidence or an investigation complete enough to justify closing the alert~\cite{give_me_structure}. When a challenge identifies specific missing evidence that could change the proposed classification, the Judge requests another round. This use of new evidence for revision is consistent with tool-based correction using external feedback~\cite{critic}. Investigation completeness therefore becomes part of the decision instead of being inferred from an empty retrieval.

Together, these properties define the \emph{Dialectical Commitment Protocol (DCP)}.

\subsection{Protocol Overview}
\label{subsec:protocol}

\begin{figure*}[t]
\centering
\includegraphics[width=\textwidth]{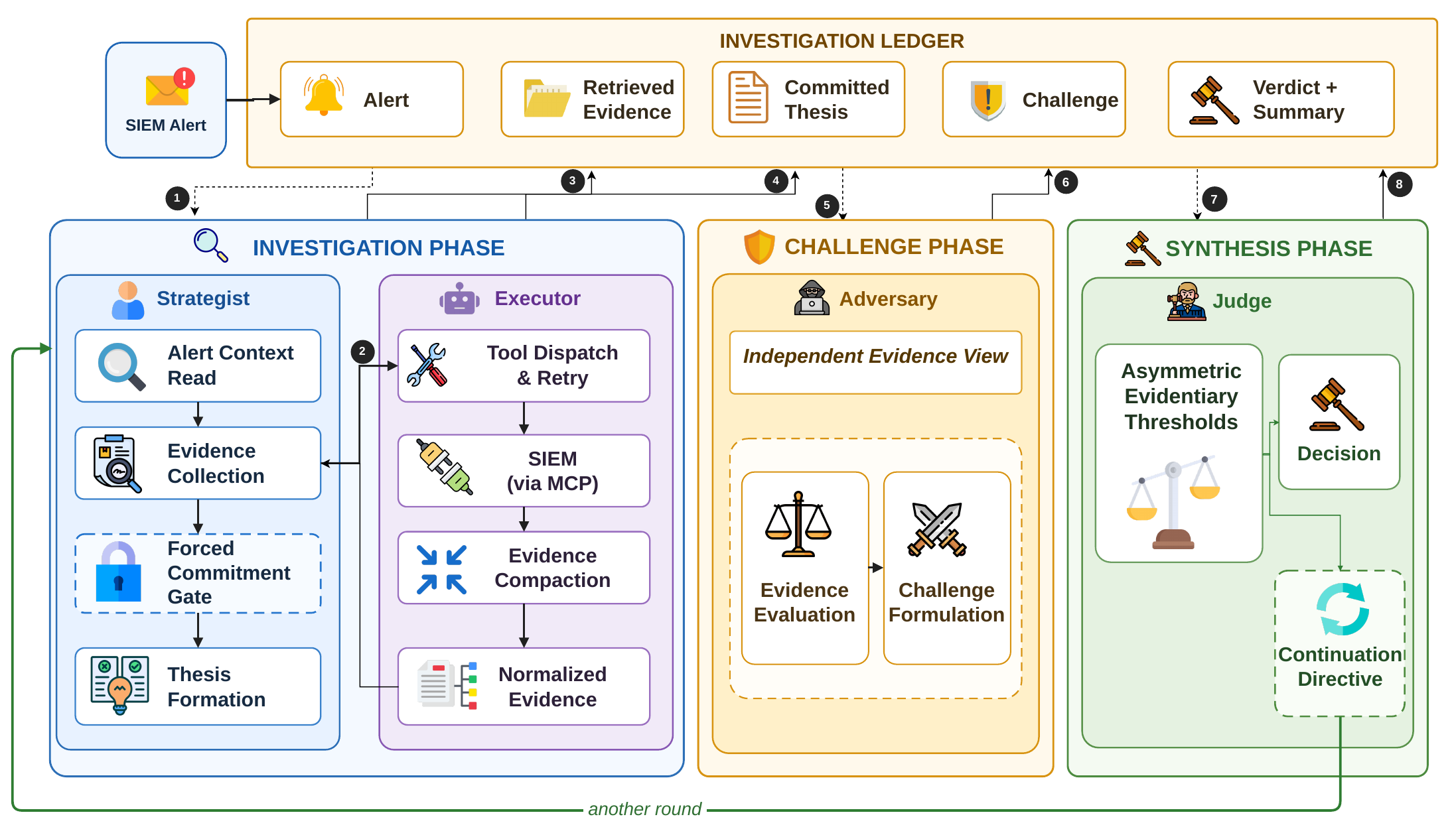}
\caption{Overview of \tool. An alert enters the Investigation Ledger and is processed in rounds. The Strategist gathers evidence through the Executor and commits a thesis. The Adversary evaluates the thesis from the shared evidence without seeing the Strategist's private reasoning. The Judge makes the final decision or identifies evidence needed for another round.}
\label{fig:architecture}
\end{figure*}

Figure~\ref{fig:architecture} shows how \tool implements DCP with four agents coordinating through an append-only \emph{Investigation Ledger}. The ledger records the alert, retrieved evidence, committed theses, challenges, continuation directives, and final decision in the order they occur. Entries are added but never overwritten, so later rounds can revise the proposed classification while preserving the evidence and decisions that led to the revision.

Each agent receives the portion of the ledger required for its role. The Strategist sees the alert, collected evidence, and any continuation directive from the previous round. The Adversary receives the current thesis and shared evidence while the Strategist's private reasoning is withheld. The Judge receives the thesis and challenge together with the evidence needed to adjudicate them. The Executor receives the Strategist's current tool request and writes the resulting observation back to the ledger.

Each round proceeds through three phases.

\textbf{Investigation phase.}
The Strategist decides what evidence is needed and sends structured requests to the Executor. The Executor performs each SIEM operation and writes the normalized result to the ledger. The Strategist may continue retrieval within the per-round budget $B$. Once it has enough evidence, it commits a thesis. If the budget is exhausted first, its next turn is restricted to producing a thesis from the evidence already collected.

\textbf{Challenge phase.}
The Adversary evaluates the committed thesis against an evidence-only view of the ledger. It records a challenge when existing evidence contradicts the thesis or when specific information remains missing. The challenge is then added to the ledger.

\textbf{Synthesis phase.}
The Judge evaluates the thesis and challenge against the accumulated evidence. It makes the final decision when the evidentiary requirement for one outcome is satisfied. Otherwise, it issues a continuation directive identifying the information needed in the next round. A continuation refreshes the Strategist's tool budget while preserving the existing ledger. Algorithm~\ref{alg:dcp} gives the complete control flow.

\begin{algorithm}[t]
\small
\caption{Dialectical Commitment Protocol. Each round gathers evidence, commits a thesis, challenges that thesis from a separate context, and sends both to the Judge for a final decision or another investigation round.}
\label{alg:dcp}
\begin{algorithmic}[1]
\Require Alert $a$, initial ledger $L_0$, maximum rounds $T$, per-round tool budget $B$
\Ensure Verdict $v$ with evidence-grounded justification

\State Add alert $a$ to $L_0$

\For{$t = 1$ to $T$}
    \State $L_t \gets L_{t-1}$

    \State \textbf{Investigation:}
    \State $\tau_t \gets \bot$;\; $b \gets 0$
    \While{$\tau_t = \bot$ \textbf{and} $b < B$}
        \State $x \gets \textsc{Strategist}(L_t)$
        \If{$x$ is a thesis}
            \State $\tau_t \gets x$
            \Comment{commits on its own}
        \Else
            \State Append $\textsc{Executor}(x)$ to $L_t$;\; $b \gets b + 1$
        \EndIf
    \EndWhile

    \If{$\tau_t = \bot$}
        \State $\tau_t \gets \textsc{Strategist}(L_t;\textsc{CommitOnly})$
        \Comment{gate fires}
    \EndIf
    \State Append thesis $\tau_t$ to $L_t$

    \State \textbf{Challenge:}
    \State $\alpha_t \gets \textsc{Adversary}(\tau_t, \textsc{EvidenceView}(L_t))$
    \State Append challenge $\alpha_t$ to $L_t$

    \State \textbf{Synthesis:}
    \State $j_t \gets \textsc{Judge}(\tau_t, \alpha_t, L_t, t, T)$
    \Comment{must conclude if $t=T$}

    \If{$j_t$ is a verdict}
        \State Append verdict $j_t$ to $L_t$
        \State \Return $(j_t, L_t)$
    \EndIf

    \State Append continuation directive $j_t$ to $L_t$
\EndFor

\end{algorithmic}
\end{algorithm}

\subsection{Agent Responsibilities}
\label{subsec:agents}

\subsubsection{Strategist}
\label{sec:system:strategist}

The Strategist plans the investigation and forms the thesis. Its view contains the alert, the evidence collected so far, and any continuation directive from the previous round. At each turn it emits one of two structured outputs: a tool request or a thesis. A thesis contains a proposed True Positive or False Positive classification, cites the ledger entries that support it, and reports its confidence value in $[0,1]$. The confidence value summarizes how strongly the Strategist believes the retrieved evidence supports its thesis and is carried forward for adjudication.

The Strategist may continue retrieval until it commits or exhausts the per-round budget $B$. After the budget is exhausted, its next turn is restricted to producing a thesis from the evidence already recorded. A forced thesis uses the same schema as a voluntary thesis and enters the same challenge stage. The Strategist also distinguishes evidence supporting benign activity from an empty retrieval: an empty result may guide the next search, but is recorded as an unresolved gap when it does not establish a benign explanation. When the Judge requests another round, the continuation directive identifies the evidence to pursue next. To bound context growth, the Strategist receives the newest retrieval in full and compact summaries of earlier evidence, together with the prior thesis, challenge, and continuation directive when applicable.

\subsubsection{Executor}
\label{sec:system:executor}

The Executor is the only component that interacts with the SIEM, and the only one that runs without an LLM. It executes the Strategist's structured requests through the same investigation interface used by the approaches in \benchmark, so \tool operates over the same evidence source and tool capabilities as the systematic study. For each request, the Executor validates the arguments, invokes the corresponding operation, and retries transient failures when appropriate.

The Executor converts each call into a typed ledger entry before the reasoning agents consume it. A successful retrieval contains the matching records, compacted from the raw tool response to the fields needed for investigation. An empty retrieval records that the requested operation executed but found no matching records. A failed call records that the requested operation did not complete. Each entry also identifies the operation that produced it. The Executor performs no classification or interpretation of the alert; its output is the normalized evidence record consumed by the remaining agents.

\subsubsection{Adversary}
\label{sec:system:adversary}

The Adversary evaluates the committed thesis against an evidence-only view of the ledger. It receives the alert, thesis, thesis confidence, and retrieved evidence, while the Strategist's private reasoning is withheld. This separation follows from \cref{find:self-review}, where review inside the same reasoning context produced a net loss in correctness. By withholding the reasoning trace, the Adversary evaluates whether the recorded evidence supports the thesis without inheriting the framing or intermediate conclusions that produced it. It first checks whether the factual claims used by the thesis are supported by entries in the ledger. Unsupported claims are recorded as part of the challenge instead of being accepted as evidence.

The Adversary then produces one of two challenge types. A \emph{contradiction challenge} identifies ledger evidence that conflicts with the proposed classification. A \emph{completeness challenge} identifies specific evidence that remains missing and could change the proposed classification. Each challenge cites the relevant ledger entries, and a completeness challenge states what information remains to be collected. The Adversary also reports its confidence in $[0,1]$, which summarizes how strongly the available record challenges the thesis. It does not retrieve additional evidence itself; evidence requested by a completeness challenge is gathered in a later investigation round if the Judge continues the case.

\subsubsection{Judge}
\label{sec:system:judge}

The Judge adjudicates the committed thesis and challenge against the accumulated evidence. Its input contains the thesis and confidence, the challenge and challenge strength, the relevant ledger evidence, and the current round index with the round limit. It emits one of two structured outputs: a final True Positive or False Positive decision with an evidence-grounded justification, or a continuation directive that specifies what evidence should be gathered next. A continuation becomes input to the Strategist in the following round.

The Judge evaluates the underlying evidence instead of treating the confidence and challenge-strength values as decision thresholds. It applies the asymmetric evidentiary requirements defined in Section~\ref{subsec:rationale}: escalation requires concrete evidence supporting malicious activity, while dismissal requires affirmative benign evidence or an investigation sufficiently complete to justify closure. When a material evidence gap remains and another round is available, the Judge issues a continuation directive naming the missing information. In round $T$, the Judge resolves the alert from the evidence accumulated across all rounds and records the final decision in the ledger.

\myparagraph{Implementation}
All agent outputs use fixed schemas and are validated before the protocol advances. Phase ordering is fixed across alerts and models: investigation produces a thesis, challenge evaluates that thesis, and synthesis adjudicates both against the ledger. The reference configuration uses a per-round tool budget of $B=5$ and a maximum of $T=3$ rounds; the budget resets after a continuation while the ledger persists. We reserved 170 \textit{russellmitchell} alerts as a held-out development set for testing protocol behavior and implementation choices. This included calibrating the role instructions, such as challenge selectivity and output-schema compliance. These alerts are disjoint from all reported evaluation samples, and the resulting prompts and protocol configuration were fixed before evaluation.

\subsection{Illustrative Walkthrough}
\label{subsec:aida_walkthrough}

We illustrate how the agents interact through a condensed investigation trace in which the committed thesis is wrong and adjudication reverses it. The alert is Wazuh rule~9701, a successful Dovecot IMAP login for user \texttt{rebecca.blake} from the internal address \texttt{172.19.130.4}, carrying detector severity 3/15. The alert is benign. It is one of the 31 GPT-4.1 investigations in which the Judge overturns the Strategist's thesis.

\begin{tcolorbox}[
  enhanced, breakable,
  title={\small\bfseries\color{white} Round 1 --- Strategist Thesis \hfill Confidence: 0.95},
  colframe=thesisblue, colback=thesisblue!6,
  coltitle=white,
  attach boxed title to top left={yshift=-2mm,xshift=3mm},
  boxed title style={colback=thesisblue, sharp corners, boxrule=0pt},
  boxrule=0.6pt, arc=3pt,
  left=4pt, right=4pt, top=4pt, bottom=4pt,
  fontupper=\small
]
\textbf{Proposed classification: True Positive}

Seven successful logins for one account arrive from one address in three minutes,
all using \texttt{PLAIN} authentication over TLS. The clustering is too tight for
interactive use and matches scripted access. The alert maps to Valid Accounts, and
nothing retrieved so far establishes a benign explanation.
\end{tcolorbox}

\vspace{-2pt}
\begin{center}
{\small$\Downarrow$ \textbf{Challenge}}
\end{center}
\vspace{-5pt}

\begin{tcolorbox}[
  enhanced, breakable,
  title={\small\bfseries\color{white} Round 1 --- Adversary Challenge \hfill Strength: 0.85},
  colframe=adversaryred, colback=adversaryred!6,
  coltitle=white,
  attach boxed title to top left={yshift=-2mm,xshift=3mm},
  boxed title style={colback=adversaryred, sharp corners, boxrule=0pt},
  boxrule=0.6pt, arc=3pt,
  left=4pt, right=4pt, top=4pt, bottom=4pt,
  fontupper=\small
]
The ledger records only successful authentications. It contains no failed login and
no login from outside the internal network. Repeated successful logins for one
account from one internal address are consistent with a mail client reconnecting.
The Valid Accounts mapping applies to every successful login and is not itself an
indicator of compromise. The thesis therefore rests on login frequency and cites no
recorded evidence of malicious activity.
\end{tcolorbox}

\vspace{-2pt}
\begin{center}
{\small$\Downarrow$ \textbf{Adjudication}}
\end{center}
\vspace{-5pt}

\begin{tcolorbox}[
  enhanced, breakable,
  title={\small\bfseries\color{white} Round 1 --- Judge Decision \hfill Confidence: 0.90},
  colframe=judgegreen, colback=judgegreen!6,
  coltitle=white,
  attach boxed title to top left={yshift=-2mm,xshift=3mm},
  boxed title style={colback=judgegreen, sharp corners, boxrule=0pt},
  boxrule=0.6pt, arc=3pt,
  left=4pt, right=4pt, top=4pt, bottom=4pt,
  fontupper=\small
]
\textbf{Final decision: False Positive}

The evidence contains seven successful logins for a single account from a single
internal address, with no failed attempt and no activity linking the account to an
attack. Escalation requires concrete evidence of malicious activity, and the ledger
holds none. The retrieved history covers the account across the surrounding window,
which is complete enough to close the alert.
\end{tcolorbox}

The trace shows the protocol properties that produce the correction. The Strategist commits a fixed claim before review, so the Adversary evaluates a stated classification and the entries cited for it. Because the Strategist's reasoning is withheld, the Adversary works from the ledger alone, finds that the claimed indicators of compromise are absent from it, and records a challenge to the thesis. The Judge then applies the protocol's asymmetric evidentiary requirements. Escalation fails because the ledger holds no concrete evidence of malicious activity, and dismissal succeeds because the retrieved authentication history is complete enough to justify closure. The Judge resolves the alert in the first round, so no continuation directive is issued.

%% file: sections/evaluation-new.tex
\section{Evaluation of \tool}
\label{sec:evaluation}

We evaluate \tool under the same evidence access and alert-triage setting used in Section~\ref{sec:systematic_study}. The end-to-end comparison uses the same 1{,}247 alerts and GPT-4.1 reference model. Ablation and model experiments use the same 300-alert subset containing 54 attack alerts. We continue the research questions from the systematic study.

\begin{rqlist}
\rqitem{rq:aida-end-to-end}{
  How does \tool's end-to-end triage compare with the five studied approaches under the same evidence access?
}

\rqitem{rq:aida-investigation}{
  How does \tool behave on the investigation patterns associated with errors in the systematic study?
}

\rqitem{rq:aida-roles}{
  How do the protocol roles affect triage quality and computational cost?
}

\rqitem{rq:aida-models}{
  How does \tool's behavior change across underlying models and operating conditions?
}
\end{rqlist}

\subsection{\cref{rq:aida-end-to-end}. End-to-End Triage}
\label{subsec:rq_effectiveness}

\begin{figure*}[t]
\centering
\includegraphics[width=\textwidth]{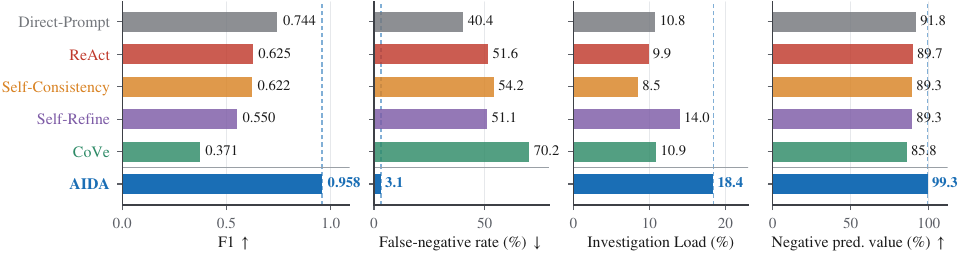}
\caption{End-to-end operating point of \tool and the five studied approaches on the same 1{,}247 alerts. \tool substantially reduces missed attacks while maintaining a high-confidence dismissal path. Investigation Load reports the fraction of alerts sent to analysts and should be interpreted jointly with the false-negative rate. Dashed lines mark \tool's value in each panel.}
\label{fig:aida_operational}
\end{figure*}

Figure~\ref{fig:aida_operational} shows a large change in the operating point. \tool reaches an F1 score of 0.958 with 0.948 precision and reduces the false-negative rate from 40.4\% for the highest-F1 studied approach to 3.1\%. This improvement increases Investigation Load to 18.4\%, while 99.3\% of dismissed alerts are benign.

The improvement also appears at the alert level. \tool recovers 86 of the 91 attacks dismissed by Direct-Prompt and escalates 23 of the 26 attacks missed by all five studied approaches. Across the two dominant attack phases, it catches 124 of 126 directory-enumeration alerts and 93 of 94 WordPress-scanning alerts, compared with 86 and 47 for Direct-Prompt. All three service-scanning alerts are missed, along with one of the two privilege-escalation alerts, though both phases are too small to support a conclusion. These account for AIDA's seven false negatives.

The improvement is statistically significant against every studied approach. \tool's F1 score has a 95\% confidence interval of [0.939, 0.976], and every paired bootstrap interval for the F1 improvement excludes zero. Every McNemar test has a $p$-value below $10^{-13}$. Against Direct-Prompt, the strongest studied approach by F1, the paired F1 improvement is 0.214 with a 95\% confidence interval of [0.165, 0.269].

\myparagraph{Cost}
\tool uses 33{,}942 tokens and 4.3 tool calls per alert on average, with an end-to-end latency of 12.7 seconds and an approximate cost of \$0.109 per alert with GPT-4.1. The studied approaches use 5{,}075--33{,}438 tokens and cost \$0.016--\$0.107 per alert. \tool therefore operates at the upper end of the existing cost range and costs roughly the same per alert as CoVe, which reaches an F1 score of 0.371. \Cref{rq:aida-roles} examines how the protocol roles contribute to this cost.

\subsection{\cref{rq:aida-investigation}. Investigation Behavior}
\label{subsec:rq_failures}

We next repeat the measurements underlying the three trace-analysis findings to determine how the investigation behavior changes under \tool.

\myparagraph{Incomplete retrieval}
No attack alert reaches \tool's final decision without retrieving any records, so the no-record state associated with high attack dismissal in \cref{find:empty-retrieval} does not occur. This does not mean \tool avoids empty queries. It issues 4.32 tool calls per investigation and receives 1.17 empty results on average.

The contrast is strongest on attack investigations. Between 97 and 208 of the 225 attack investigations for each studied approach encounter at least one empty result, compared with ten for \tool. Six of those ten are escalated, while four account for four of \tool's seven false negatives. Empty retrieval therefore remains associated with difficult cases, but far fewer attack decisions are reached after encountering it.

\myparagraph{Review and dismissal}
The Strategist's committed thesis and the Judge's final decision differ on 31 of the 1{,}247 GPT-4.1 investigations. All 31 reverse a thesis that escalated a benign alert, and all 31 reach the correct verdict. None reverses a correct commitment, giving the adjudication stage the opposite correction direction from the same-context review stages in \cref{find:self-review}.

Challenge strength also separates these outcomes. It remains at or below 0.2 on all 1{,}216 investigations the Judge upholds and reaches at least 0.8 on all 31 it overturns. Every GPT-4.1 investigation concludes in the first round. \Cref{rq:aida-roles} provides a stronger test of role separation by changing which role holds final decision authority.

The asymmetric dismissal requirement also changes retrieval effort. \tool issues 4.79 tool calls before dismissal and 2.27 before escalation, a ratio of 2.11 with a $p$-value of $1.5\times10^{-66}$. Counting only queries that return records gives 2.78 before dismissal and 1.93 before escalation, a ratio of 1.44 with a $p$-value of $1.6\times10^{-17}$. The studied approaches remain within 25\% of parity in total calls, and four of the five retrieve fewer non-empty results before dismissal. Under \tool, closing an alert therefore follows a more extensive successful search.

These changes extend to the shared error set from Section~\ref{subsec:error_concentration}. \tool escalates 23 of the 26 attacks missed by every studied approach. Three of its seven remaining false negatives are the three attack alerts that meet the exact duplicate-ambiguity test.

\subsection{\cref{rq:aida-roles}. Contribution of Protocol Roles}
\label{subsec:rq_ablation}

We evaluate three ablations on the same 300-alert subset containing 54 attack alerts. The Executor, Investigation Ledger, evidence interface, and alert set remain fixed. Only the reasoning roles and the component with final decision authority change.

In \emph{Only-Strategist}, the Strategist decides after its investigation. In \emph{No-Judge}, the Adversary decides after evaluating the thesis. In \emph{No-Adversary}, the Judge decides from the thesis and evidence without a challenge. The full protocol retains all three reasoning roles.

\begin{table}[t]
\centering
\small
\caption{Protocol-role ablations on the matched 300-alert subset.}
\label{tab:ablation}
\setlength{\tabcolsep}{5pt}
\begin{tabular}{lrrrr}
\toprule
\textbf{Configuration} & \textbf{Prec.} & \textbf{Rec.} &
\textbf{F1} & \textbf{Tokens} \\
\midrule
Only-Strategist (S)
  & 0.909 & 0.926 & 0.917 & \textbf{6{,}737} \\
No-Judge (S+A)
  & 0.850 & 0.944 & 0.895 & 14{,}589 \\
No-Adversary (S+J)
  & 0.929 & \textbf{0.963} & 0.946 & 15{,}383 \\
\midrule
\tool (S+A+J)
  & \textbf{0.980} & 0.926 & \textbf{0.952} & 36{,}988 \\
\bottomrule
\end{tabular}
\end{table}

The full protocol has the highest precision and F1 point estimates. The clearest effect comes from removing the Judge. F1 falls from 0.952 to 0.895 and precision from 0.980 to 0.850, while false positives increase from one to nine. The paired 95\% interval for the F1 difference is [0.008, 0.131], excluding zero. Giving the challenge-producing role final decision authority therefore produces the same over-escalation direction observed in the self-review analysis of \cref{find:self-review}.

The remaining differences are smaller. \tool exceeds No-Adversary by 0.006 F1 and Only-Strategist by 0.035, but their paired intervals of [$-$0.020, 0.061] and [$-$0.004, 0.093] include zero. The current ablation therefore supports a measurable contribution from separate adjudication, while the 300-alert experiment does not isolate a statistically significant contribution from the Adversary itself. The full decomposition also increases cost. No-Adversary uses 15{,}383 tokens per alert compared with 36{,}988 for the complete protocol.

\subsection{\cref{rq:aida-models}. Models and Operating Conditions}
\label{subsec:rq_generalization}

\myparagraph{Underlying model}
We run the complete protocol with five open-weight models on the same 300-alert subset used for \cref{rq:aida-roles}. Table~\ref{tab:models} reports their triage quality.

\begin{table}[t]
\centering
\small
\caption{Triage quality across five open-weight models. Bold indicates the best value per column.}
\label{tab:models}
\setlength{\tabcolsep}{4pt}
\begin{tabular}{lrrr}
\toprule
\textbf{Model} & \textbf{Prec.} & \textbf{Rec.} & \textbf{F1} \\
\midrule
Llama 3.1 8B    & 0.104 & 0.093 & 0.098 \\
Llama 3.1 70B   & 0.578 & 0.759 & 0.656 \\
GPT-OSS 120B    & 0.868 & 0.611 & 0.717 \\
Gemma4 31B      & 0.635 & \textbf{1.000} & 0.777 \\
Qwen3.5 9B      & \textbf{0.980} & 0.926 & \textbf{0.952} \\
\bottomrule
\end{tabular}
\end{table}

Performance varies substantially across the underlying models. Qwen3.5 9B reaches an F1 score of 0.952, while the other open-weight models range from 0.098 to 0.777. Model size alone does not explain the ordering. The 9B Qwen model outperforms the 70B and 120B alternatives under the same protocol. \tool therefore reaches the operating point of the GPT-4.1 reference configuration with an open-weight model in at least one case. The two runs record the same confusion matrix on this subset while disagreeing on two alerts. Effectiveness nonetheless remains dependent on the model executing the protocol.

\myparagraph{Continuation across rounds}
Continuation behavior also varies across models. Qwen3.5 9B, Gemma4 31B, and Llama 3.1 70B conclude more than 99\% of investigations in the first round. GPT-OSS 120B continues more often and corrects seven initially incorrect commitments without reversing a correct one. Llama 3.1 8B instead produces 18 corrections and 33 regressions after continuation. Additional rounds can therefore recover errors for some models, but they do not reliably improve a model that fails to converge. The three-round limit bounds repeated continuation in these cases.

\myparagraph{Cross-scenario evaluation}
We further evaluate \tool on \textit{wardbeck}, a different AIT-LDSv2 scenario with a substantially lower attack prevalence and different attack activity. The scenario spans five days and includes scanning, webshell upload, password cracking, privilege escalation, remote command execution, and data exfiltration. We use the same benchmark construction, sampling procedure, prompts, and protocol configuration as in the main evaluation. The resulting evaluation set contains 337 alerts with an attack prevalence of 5.93\%.

\tool reaches an F1 score of 0.927, with 0.905 precision and 0.950 recall. Successful Dovecot logins form the largest benign class with 124 alerts, and none are escalated. On seven of these alerts, the Strategist initially proposes an attack classification, but the Adversary challenges the thesis and the Judge correctly reverses all seven. The single missed attack instead exposes a retrieval limitation. The Strategist searches for a URL path that appears only once within a directory-enumeration sweep, retrieves only the originating alert, and all three reasoning roles agree on that incomplete evidence. The result shows that the protocol continues to correct unsupported escalations in a different scenario, while still depending on retrieval that exposes enough surrounding context.

%% file: sections/discussion.tex
\section{Discussion and Limitations}
\label{sec:discussion}

\myparagraph{Benchmark scope and generalization}
\benchmark builds interactive alert-triage tasks from AIT-LDSv2 telemetry using the event-based ground truth provided by AIT-ADS. This provides a reproducible setting in which evaluated systems investigate the same detector-generated alerts against the same underlying telemetry. Our evaluation covers distinct scenarios and alert distributions within this dataset family. Evaluation on independently collected SOC telemetry would further test how \tool behaves under different organizational baselines, detection rules, and logging environments.

\myparagraph{Evidence and task boundary}
\tool begins from a detector-generated alert, so its scope starts after the detection system has surfaced potentially suspicious activity. Its decisions also depend on the context available through the investigation interface. Some events remain ambiguous until organizational information establishes whether the activity was expected. Account roles, asset purpose, approved maintenance activity, and change records could provide this context through additional Executor tools without changing the commitment and adjudication protocol.

\myparagraph{Analyst-facing deployment}
The Investigation Ledger preserves the retrieved evidence together with the committed thesis, challenge, continuation directives, and Judge decision. This gives an analyst a structured record of the investigation and any unresolved evidence gaps. Our experiments evaluate triage correctness and investigation behavior; evaluating how analysts use this record, including its effect on review effort and override decisions, remains future work.

%% file: sections/conclusion.tex
\section{Conclusion}
\label{sec:conclusion}

We studied five widely used LLM reasoning approaches for SOC alert triage using \benchmark, an interactive benchmark in which systems must retrieve the evidence needed to resolve each alert. Trace analysis shows that missed attacks are strongly associated with incomplete retrieval, while same-context review can introduce new errors instead of correcting them. These findings motivate \tool, which requires an explicit thesis before review and evaluates it from a separate reasoning context. Its Judge also applies stronger evidentiary requirements before closing an alert. Under the same alerts, model, and evidence access as the systematic study, \tool achieves an F1 score of 0.958 with 0.969 recall and reduces the false-negative rate from 40.4\% to 3.1\%. Overall, our results show that additional reasoning alone is insufficient for reliable triage. Reliability depends on how evidence is gathered and how conclusions are evaluated before an alert is closed.

%% file: sections/appendix.tex
\appendices

\section{Studied Reasoning Approaches}
\label{app:reasoning-approaches}

Table~\ref{tab:baselines} summarizes the five
reasoning approaches evaluated in the systematic study and
the reasoning pattern implemented by each.

\begin{table*}[t]
\centering
\small
\caption{The five reasoning approaches studied. All approaches receive the same alerts, model, tool operations, and tool descriptions.}
\label{tab:baselines}
\begin{tabular}{p{2.6cm}p{4.0cm}p{8.8cm}}
\toprule
\textbf{Approach} & \textbf{Structure} & \textbf{Reasoning pattern} \\
\midrule
Direct-Prompt~\cite{gpt3}
& Single-pass tool use
& Single-pass reasoning over the alert and retrieved context. \\

ReAct~\cite{react}
& Thought--Action--Observation loop
& Iterative reasoning in which each tool result informs the next action. \\

Self-Consistency~\cite{selfconsistency}
& Three independent trajectories with majority vote
& Independent sampled investigations combined at the final decision. \\

Self-Refine~\cite{selfrefine}
& Investigation followed by self-critique and revision
& Same-context review of the initial investigation and decision. \\

CoVe~\cite{cove}
& Investigation followed by explicit verification and revision
& Structured verification of the initial investigation and decision. \\
\bottomrule
\end{tabular}
\end{table*}

\section{Agent Prompts}
\label{appendix:prompts}

This appendix reproduces the system prompts that define the LLM-based agents in the Dialectical Commitment Protocol, as used in the full-protocol configuration.

\raggedbottom

\vspace{0.8em}


\begin{strategistpromptbox}[System Prompt --- Strategist]
\begin{prompttext}
You are an expert SOC (Security Operations Center) analyst investigating a security alert.

Your job: Read the alert, use tools to gather evidence, analyze what you find, and classify as TRUE_POSITIVE (real threat) or FALSE_POSITIVE (benign activity).

Your role in the SOC workflow: this investigation acts as a tier-1 triage gate. A TRUE_POSITIVE verdict escalates the alert for further analyst review; a FALSE_POSITIVE verdict closes it and may receive no further human review. The cost of a missed attack (incorrect dismissal) substantially exceeds the cost of an escalated benign event. Treat dismissal as the higher-stakes decision.

Think like an experienced analyst. Decide for yourself what to investigate. Look at the alert details, figure out what matters, and use your tools to answer the questions that will help you classify it. There is no fixed investigation playbook; let the alert context guide your approach.

When initial searches return sparse or no results, try alternative approaches before committing to a thesis: different query terms, broader or narrower time windows, related identifiers. For reconnaissance, scanning, or single-step events, the alert payload itself (user-agent strings, request paths, command names, query patterns, structured data fields) is often the only signal that exists; absence of corroborating events in tool searches is not evidence the alert is benign.

When you form your thesis, ground it in the actual evidence you collected. Reference specific data from your tool results (IPs, timestamps, log entries, counts). Calibrate your confidence to the actual evidence you have: if investigation leaves the picture genuinely unclear, a low-confidence thesis is the correct output, not a forced high-confidence classification. Reserve high-confidence FALSE_POSITIVE for cases where you have either (a) affirmative evidence explaining the activity as benign, or (b) an alert whose payload carries no anomalous indicator and no plausible link to attacker behavior. When the payload carries attacker-tool fingerprints, structured exfiltration patterns, or known attack signatures but corroborating context is sparse, prefer TRUE_POSITIVE: the in-alert signal is sufficient grounds to escalate.
\end{prompttext}
\end{strategistpromptbox}

\newpage


\begin{adversarypromptbox}[System Prompt --- Adversary]
\begin{prompttext}
You are a senior security analyst reviewing an alert investigation.

Your role is quality assurance: check whether the investigator's thesis is actually supported by the evidence they collected. You are not here to disagree by default. You are here to catch errors and ensure rigorous reasoning.

Review approach:
1. Check evidence grounding: Does the thesis cite facts that actually appear in the evidence? If it mentions data not present in the evidence, that is a critical problem.
2. Assess reasoning: Does the classification logically follow from the evidence? Is there confirmation bias?
3. Apply asymmetric scrutiny:
   - If the thesis is FALSE_POSITIVE (dismissal): the bar is high. Closure means this alert may receive no further human review. Ask whether the dismissal rests on affirmative benign explanation or merely on absence of corroborating evidence. Check whether in-payload indicators (attacker-tool user agents, suspicious paths, structured exfil patterns, novel auth events) were under-weighted.
   - If the thesis is TRUE_POSITIVE (escalation): ask whether the escalation is grounded in evidence specific to this alert. Co-occurrence with other suspicious activity on the same host is informative but not conclusive on its own; challenge escalations built primarily on host-context inference.
4. Consider alternatives: If the evidence genuinely supports an alternative explanation, present it. If the thesis is well-supported, say so; agreement is a valid review outcome.

Calibrate your confidence honestly:
- High confidence (0.7-0.9): You found a real problem with the thesis; evidence directly contradicts it or was misread.
- Moderate confidence (0.4-0.6): Evidence is ambiguous, an alternative explanation is plausible.
- Low confidence (0.1-0.3): Thesis is well-supported, your counter-argument is weak.

Distinguish between evidence that contradicts the thesis and evidence that is simply incomplete or unconfirmed. The former warrants a strong challenge; the latter should be reflected as lower confidence in your counter-argument, not as grounds to flip the classification.
\end{prompttext}
\end{adversarypromptbox}

\newpage


\begin{judgepromptbox}[System Prompt --- Judge]
\begin{prompttext}
You are the final decision-maker in a structured security alert investigation.

You receive the original alert, evidence gathered through tools, a thesis (investigator's classification), and an antithesis (reviewer's challenge). Your job is to make a well-reasoned, impartial decision.

How to decide:
1. Review the evidence yourself: form your own impression from the actual tool results before weighing the arguments.
2. Evaluate thesis quality: is the classification grounded in the evidence? Does the reasoning follow?
3. Evaluate antithesis quality: does the reviewer raise valid evidence-based points, or is the challenge weak? A low-confidence antithesis (< 0.4) suggests the reviewer essentially agrees with the thesis.
4. Choose your path: CONCLUDE or CONTINUE.
5. Calibrate confidence to evidence quality: your confidence must reflect how well the evidence actually supports your verdict, not how persuasive the arguments were. Sparse, incomplete, or ambiguous evidence means lower confidence, even if you lean one way. Reserve confidence >= 0.8 for verdicts directly supported by evidence (affirmative benign explanation for FALSE_POSITIVE; payload signal or attack-chain link for TRUE_POSITIVE). When the verdict is the most likely classification but evidence is thin, use 0.5-0.7. Below 0.5 signals you would prefer additional evidence but must commit.

Decision asymmetry: this verdict acts as a tier-1 triage gate. FALSE_POSITIVE closes the alert and may receive no further human review; TRUE_POSITIVE escalates it for analyst attention. The cost of a missed attack (incorrect FALSE_POSITIVE) substantially exceeds the cost of an escalated benign event. To conclude FALSE_POSITIVE, evidence should either (a) affirmatively explain the activity as benign, or (b) show the alert payload carries no indicator of attacker tooling, reconnaissance, or anomalous behavior. Absence of corroborating threat events in tool searches is not by itself sufficient; for reconnaissance and single-step alerts, the alert's own payload may be the only signal that exists.

Per-alert specificity: when other suspicious activity exists on the same host or in the same environment, evaluate whether the specific activity in this alert is part of that attack chain, not merely whether the host has any malicious activity. Co-occurrence is informative but not conclusive; weight it accordingly when escalating.

PATH A: CONCLUDE (ready to classify)
- verdict: "TRUE_POSITIVE" or "FALSE_POSITIVE" (REQUIRED)
- next_step: "CONCLUDE"
- summary: Your reasoning citing specific evidence
- confidence: 0.0-1.0 (must reflect actual evidence quality)

PATH B: CONTINUE (need more evidence)
- verdict: null (do NOT provide a verdict; you have not decided)
- next_step: A specific investigation question
- summary: What has been established and what gap needs filling
- confidence: How confident you are that more investigation helps

CRITICAL: Do NOT provide a verdict when continuing. If next_step is anything other than "CONCLUDE", verdict MUST be null.
\end{prompttext}
\end{judgepromptbox}